\documentclass[aps,prl,preprint,groupedaddress]{revtex4-1}
\usepackage{graphicx}
\usepackage{txfonts}
\usepackage{epstopdf}
\begin{document}

\title{Sieving hydrogen based on its high compressibility}
\author{H. Y. Chen$^{1}$, X. G. Gong$^2$, Z. F. Liu$^3$ and D. Y. Sun$^{1}$}
\email{Email address: dysun@phy.ecnu.edu.cn}
\address{$^1$Department of Physics, East China Normal University, Shanghai 200062, China}
\address{$^2$Department of Physics, Fudan University, Shanghai-200062, China }
\address{$^3$Department of Chemistry and Center for Scientific Modeling and Computation, Chinese University of Hong Kong Shatin, Hong Kong, China }
\date{\today}

\begin{abstract}

A molecular sieve for hydrogen is presented based on a carbon
nanotube intramolecular junction and a $C_{60}$. The small
interspace formed between $C_{60}$ and junction provides a size changeable
channel for the permselectivity of hydrogen while blocking $Ne$ and
$Ar$. The sieving mechanism is due to the high compressibility of
hydrogen.
\end{abstract}


\maketitle

With growing demand on clean energy, the production and separation
of  hydrogen with great amount represent a topic of intense
interests. One of the straightforward and commercial ways to
separate hydrogen is to use a specifically designed molecular sieve
with high selectivity and permeability. Efforts on using the
traditional molecular sieve, such as polymeric membrane, have been
ongoing for some time and the progress has been made along this
direction.\cite{x1,x2} However, the major problem related to
polymeric membranes is the strong trade-off between
selectivity and permeability.\cite{x1,x2} Thus, the search of new
sieves or new methods to improve the permselectivity of hydrogen
from mixtures is much needed to match the growing demand in the
hydrogen energy.\cite{y1,y2}

In molecular sieves, the major mechanism for the filtration usually
bases on the difference in size, shape or polarity of varies
molecules in the mixture. $H_2$ has almost spherical shape and no
polarity, thus the probable sieving mechanism for $H_2$ is
limited to use its small size. Attempts to design sieves based on
the size difference have been made over the past few
years.\cite{y1,y2}  This kind of sieve is quite useful for molecules
with large size difference. However, for molecules having comparable
sizes with $H_2$, the efficient way for the separation is still lack
of.

The purpose of this letter is to find an alternative mechanism to
design sieves for hydrogen. We notice that  hydrogen has a novel
property, namely the highly compressible character, which is absent
for most other molecules.\cite{y3,y4} A hydrogen molecule has only
two electrons on fairly compact 1s orbitals, the van der Waals
repulsion between two $H_2$ molecules is much weaker than typical,
so that it is highly compressible all the way up to and above GPa
pressure. The soft repulsion is also found for $H_2$ interacting
with other systems, such as graphite.\cite{26} The highly
compressible character makes hydrogen have much more chances to diffuse
at confinement space over other gases. Furthermore,
$H_2$ is the lightest molecule in nature, at a given temperature, so
$H_2$ could have higher mobility in confinement space than other
gases. Considering the above two features, it is possible to design
a molecular sieve to filtrate hydrogen.

The ideal candidate for current studies
could be carbon nanotubes (CNTs) due to its tubular structures and
small pore sizes. In fact, over the past few years, many
studies
 have focused on the investigation of the selectivity and
transport properties of light gases in CNTs.\cite{3,4,5,6,x5}
However the use of CNTs as a molecular sieve for hydrogen is still a
great challenge.  If one hopes to directly use the pore of CNTs, the
CNT should have comparable pore size with $H_2$ ($\sim 3\AA$). Such
small pore size almost approaches to both theoretical predicated
limitation and the smallest CNT fabricated in
experiments.\cite{9,10,11}  Up to now, as far as we know, the
reliable design based on CNTs is still lack of.

To avoid the above mentioned limitation, we do not directly use the
channel of CNTs, but the interspace formed between $C_{60}$ and
CNTs. The inset of Fig. 1(a) presents the geometry of the molecular
sieve.  It is made of an intramolecular junction and a $C_{60}$. To
form the intramolecular junction between two tubes, the method
presented in Ref.\cite{12,13,14,21,22} is adopted. Combining the
well-developed experimental techniques for producing
junctions\cite{15,16,17} and for inserting $C_{60}$ into
CNTs\cite{18,19,20}, our design could be easily fabricated. In
current studies, the left tube is (20,0) CNT, which is larger than
$C_{60}$, while the right one is (5,5) CNT, smaller than $C_{60}$.
The interspace formed between $C_{60}$ and
CNTs could provide a nano channel for the separation
of hydrogen using its high compressibility.

To compare the permselectivity of hydrogen with other
gases, two typical gases are also considered, which
are $Ne$ and $Ar$. All the molecules/atoms have almost spherical shape and
no polarity, the major difference is their size. Among these
molecules, $Ar$ is the one with largest diameter (about 3.8{\AA});
while $H_2$ and $Ne$ have the similar sizes of about 3.3{\AA} and
3.1{\AA} respectively. Because the difference in sizes is not
prominent and $H_2$ even has larger size than $Ne$, the separation
could be very difficult, even impossible in traditional molecular
sieves. If hydrogen can be separated from
those molecules mentioned above, this design could also work for
most other types of molecules, such as $CH_4$, $N_2$ and $O_2$. It
needs to be pointed out that, all the three molecules can easily
pass both (20,0) and (5,5) CNTs.

To validate this molecular sieve, several calculations have
been carried out. First, to check the size and stability of the
narrow space formed between $C_{60}$ and CNT, we have studied the
absorption behavior of $C_{60}$.  Second, to verify if a single
$H_2$ molecule can easily pass the sieve without the highly
compressibility emerged, the diffusion barrier for one molecule
passing the sieve has been calculated. Third,  to validate the
importance of the high compressibility, a comparing study on the
transport properties for each pure gas in the sieve is carried out.
Finally,  the permselectivity of $H_2$ mixed with other gases is
also calculated. This work demonstrates that, if the highly
compressible character of hydrogen are taken into account, our
design does work as a molecular sieve for hydrogen.

We have addressed the above questions using molecular dynamic (MD)
simulation. To calculate the permselectivity and flux, the molecular
sieve with length about 36 ${\AA}$ is dipped into a reservoir, as
shown in Fig.1(b). Two plates, labeled as A and B, are placed at the
two boundaries of the reservoir perpendicular to the
molecular sieve. These plates, made of the corresponding gas
molecules, are rigid with face-centered cubic (111) planes. In all
simulations, plate B is fixed, while plate A is movable. In all
studies, the system temperature keeps at 300 K by
N$\acute{o}$se-Hoover thermostat.\cite{23} The equations of motion
are integrated by using the predictor-corrector algorithm with time
step of 1 fs. Periodic boundary conditions are only applied in
directions perpendicular to the CNTs. The pressure difference around
$C_{60}$ is defined as $\Delta P=P_{left}-P_{right}$, where
$P_{left}$ is fluidic pressure of the reservoir, and $P_{right}$
keeps zero in all simulations.
 In current studies, $\Delta P$ is
changed by placing the plate A at different
positions. For each move of plate A, the large CNT is blocked
by fixing a few atoms/molecules inside it for first 0.4 ns so that
the fluid in the reservoir is fully equilibrated. Afterwards, all
gas atoms/molecules are movable, and the system is
simulated for about 0.6 ns to determine the flux of gases.

The interatomic potential between carbon atoms is the Tersoff-type
many-body potential with parameters given by Brenner.\cite{24,25}
The weak interaction between carbon atoms in CNTs and $C_{60}$ is
modeled by the Lennard-Jones (L-J) potential, $V_{vdW}$ =
$C_{12}$/$r^{12}$-$C_{6}$/$r^{6}$ with $C_{6}$=20 $eV${\AA}$^{6}$
and $C_{12}$ = 2.48 $\times10^{4}$ $eV${\AA}$^{12}$. Interactions
between $H_2$ molecules are modeled by the Silvera-Goldman
potential, which correctly reproduce the equation of states for
$H_2$ in wide range of pressures.\cite{y3} The weak interaction
between $H_2$ and carbon is described by the recently fitted
potential by some of us,\cite{26} which reproduces the results over
a wide range of repulsive and attractive regions as calculated by
high level {\it ab-initio} methods.\cite{x6} The interactions of
$Ne$ and $Ar$ are L-J potential with the parameters taken from
reference\cite{27}. Within the pressure up to 1 GPa, the L-J
potential could accurately predicate melting curve and other
properties which are comparable to experimental results.\cite{27,28}
The interactions of C-Ne and C-Ar are also L-J potential and
calculated using Lorentz-Berthlot combination rules with parameters
of C-C taken from Ref.\cite{29}.

Without filling any gases, there is one stable adsorption site for
$C_{60}$ around the intramolecular junction. Fig.1(a) shows the
potential energy of the system as $C_{60}$ moving along axial
direction. At this stable adsorption site, $C_{60}$ is right located
at the axis center near the junction. The minimal distance between
$C_{60}$ and the junction is around 3.3 {\AA} as indicated in the
inset of Fig. 1 (a) labeled as 'interspace A'. The
interlayer distance from $C_{60}$ to tube wall is about 4.3 {\AA}
marked as 'interspace B'. The stable adsorption position is a
reflection of the good geometric match between junction and
$C_{60}$, namely, favorable interactions between pairs of C atoms on
CNTs and $C_{60}$ reach maximum. The adsorption energy is about
1.17eV lower than that for $C_{60}$ far from the junction. The
stable adsorption guarantees $C_{60}$ staying near the junction,
which is important for the current sieve. Due to the weak
interaction between $C_{60}$ and tube wall, $C_{60}$ can be moved
around the stable position in a certain range, which forms a 'size
changeable channel'.

The diffusion barriers for single molecule/atom passing the
molecular sieve are calculated, which is plotted in Fig. 2(a). We
find that, when the molecule locates at interspace A, the global
potential maximum appears. As the molecule locates at interspace B
(right above $C_{60}$), there is another local potential energy
maximum (flat), which is more obvious for $Ar$. There is a strong
correlation between the position of molecule/atom and displacement
of the mass center of $C_{60}$. Fig. 2 (b) and (c) show the
associated motion of the mass center of $C_{60}$ as the molecule
moving along the tube axial direction. As the molecule approaches to
the interspace B, $C_{60}$ is pushed backward and begins to move to
the tube wall. The displacement of $C_{60}$ produces
increasing space for the molecule. It reaches the maximum
as the gas molecule arrives interspace A. At the same time, the
potential energy reaches the maximum. After the molecule passes
interspace A, $C_{60}$ is ready to move back to its equilibrium
position. This implies that the energy barrier is mainly due to the
displacement of $C_{60}$. The larger the displacement of $C_{60}$
is, the higher the energy barrier is. For $Ar$, the displacement of
$C_{60}$ along radial/axial direction from the equilibrium position
is around $1.61\AA$/$1.71\AA$, while for $H_2$ or $Ne$, the
corresponding value is $1.35\AA$/$1.07\AA$ and $1.36\AA$/$1.08\AA$,
respectively. The diffusion barrier for $H_2$ and $Ne$ has the
similar value around 0.44 eV (0.45eV for Ne and 0.43eV for $H_2$),
which is lower than the barrier of 0.68eV for $Ar$. The
diffusion barrier suggests that, for a single molecule/atom, it is
very difficult to pass the molecular sieve at room temperature. As
we will discuss below, if the high compressibility of hydrogen is
emerged (plenty of $H_2$ molecules appear), the difference in
transport behavior among $H_2$, $Ne$ and $Ar$ will be significant.

In Fig. 3(a), we have shown the flux via $\Delta P$ for pure $H_2$,
$Ne$ and $Ar$ at 300 K separately. From this figure, one can see
that, with the increase of pressure difference up to 1 GPa, only
$H_2$ molecules are excluded out from the molecular sieve, while
other gases ($Ne$ and $Ar$) are blocked by $C_{60}$ with zero
permeation. As we have discussed that $Ne$ has almost the same
diffusion barrier as $H_2$ (see figure 2(a)). Thus, $H_2$ and $Ne$ should have comparable chance to pass the sieve if the diffusion of the single atom/molecule  dominates this process. However the permeability of
$Ne$ is much lower than that of $H_2$. The mystery can be disclosed
in terms of the high compressibility of hydrogen. It is well known
that the diffusion rate is proportional to the inverse of the square
root of molecular mass. However the mass issue is not enough to
explain the transport difference between $H_2$ and $Ne$. On the
contrary, the high compressibility of $H_2$ plays a key role.
Comparing with $Ne$, this character is not quite obvious when few
molecules appear or the pressure is low (figure 2 (a)). With the
increase of pressure, the high compressibility (both soft repulsion
of $H_2-H_2$ and $H_2-C$) becomes more and more important.

The high compressibility makes $H_2$ have more chance to accumulate
in interspace B. We have calculated the average number of molecules
located within a narrow window around interspace B. This
narrow window has the width of 4$\AA$ along the axial direction and
centered on $C_{60}$.  Fig. 3 (b) shows the average number of
molecules/atoms within this narrow window via $\Delta P$.
One can see that, the number of $H_2$ in this area is much more than
$Ne$. And the number increases with the increase of $\Delta P$. For
$Ne$, even at 1 GPa, the number is less than two, which is several
times smaller than that of $H_{2}$. No $Ar$ atom is found at
interspace B even at 1 GPa. The molecules located at the interspace
B will not only increase the chance moving to interspace A, but also
lower the diffusion barrier due to the inward motion of $C_{60}$
driven by the molecules located around interspace B. With growing
number of $H_2$ moving into interspace B, $H_2$ is more
easily to overcome the energy barrier and be excluded out.

To further verify that transport properties of hydrogen is dominated
by the high compressibility, rather than the small mass of $H_2$, we
have carried out two additional calculations, which is shown in the
inset of Fig. 3 (a). In the first run (long dashed line), the mass of
$H_2$ is artificially enlarged ten times. In this case, we found
that the flux is simply reduced according to the change of mass. In
the second run (dot-dashed line), the interaction of $H_2$-$H_2$ and
$H_2$-C is replaced by a typical L-J potential.\cite{30}  Our
calculations show that, in this case, the transport of hydrogen is
terminated. 

The permselectivity of $H_2$ mixed with $Ne$ and $Ar$ is
further studied separately. In this calculation, the number ratio
is chosen as  $H_2:M$ (M=Ne or Ar)=1:1, 3:1, 9:1. Fig. 4(a) and (b)
shows the flux of two components mixed gases as a function of
$\Delta P$. One can see that, in all studied cases, the permeation
of $H_2$ is nonzero and increase with the increase of $\Delta P$. In
all the simulations, there is no $Ar$ atom found to pass the
molecular sieve. In the mixture of $H_2$ and $Ne$, the flux of $Ne$,
as the pressure difference raised up to around 0.6 GPa, becomes
nonzero. However it is still much smaller than that for $H_2$
(Please see the Supporting Information for a animation of this
process). Detailed studies show that, the leakage of a few $Ne$
molecules is mainly 'abducted' by $H_2$. That is, by chance, some
$Ne$ are surrounded by $H_2$, these $Ne$ atoms pass the molecular
sieve associated with $H_2$. For the mixture of three types of gas
molecules ($H_2$, $Ne$ and $Ar$) with the ratio of 1:1:1, our
calculations also show that, only $H_2$ pass through the molecular
sieve when $\Delta P$ is in the range $0.2GPa\sim0.6GPa$, as shown
in Fig. 4 (c) (Please also see the Supporting Information for a
animation of this process). Similarly a few of $Ne$ also leak out
from the molecular sieve as $\Delta P$ raising up to 0.6 GPa. As
$\Delta P$ up to 1GPa, no $Ar$ penetrates the molecular sieve during
at least 1ns simulation time.

The final question, which should be addressed,  relates to the
requirements on using the high compressible
character of $H_2$.  For this purpose, we have checked the sieving function for
hydrogen in a few combinations of intramolecular junction and $C_{60}$.
Our results show that the position of $C_{60}$ and the size of the interspace formed between the tube wall and $C_{60}$ are the important factors. First, the size of interspace should not be too small, otherwise the leakage of $H_2$
is blocked. The size of interspace should also not be
too large, otherwise other gases could penetrate and the high
compressibility of $H_2$ plays less role. We found that the interspace between the tube wall and $C_{60}$ should be around 3.5$\AA$, correspondingly, the radius of the larger CNT in the sieve around $16\AA$ is the ideal one. For the smaller tube, it needs to have enough space for hydrogen passing. Second, the potential well of $C_{60}$ should not be too stiff,  otherwise $C_{60}$ could not move to produce more space. For example, the intramolecular junction formed between (20,0) and (10,0), which fills the requirements, also has the
function as a molecular sieve.  We believe that, besides of the current design, there could be many structures for using the high compressible
character and the small mass of $H_2$.

In conclusion, by taking the high compressibility of
hydrogen into account, we have proposed a potential molecular sieve
for hydrogen based on $C_{60}$ and carbon nanotube intramolecular
junction. Our calculations show that the narrow interspace formed between the tube wall and $C_{60}$ provides a size changeable channel. The diffusion barriers for a single molecule/atom passing this channel are quite high (0.45eV for Ne, 0.43eV for $H_2$, and 0.68eV for Ar). The diffusion barriers imply that, for a single molecule/atom, it  is difficult to pass the molecular sieve at room temperature. If plenty of $H_2$ appear ( the high compressibility of $H_2$ is emerged), the transport behavior of $H_2$, $Ne$ and $Ar$ is significant difference. Using molecular dynamics simulations, we have studied the permselectivity of $H_2$/Ar, $H_2$/Ne and $H_2$/Ar/Ne mixtures. Our
calculations show that  no $Ar$ is found to pass
the sieve, while $H_2$ can pass the sieve with noticeable flux.
A few $Ne$ atoms can also leak out, however
the amount is much smaller than $H_2$. The selective permeation of
$H_2$ is the result of the high compressibility of
$H_2$. The current design could be an alternative technique for the
separating of $H_2$ from most of mixtures.

\begin{acknowledgments}
{\bf Acknowledgment.} This research is supported by Key Project of Shanghai Education Committee, Shanghai Municipal Science and Technology Commission, the National Science Foundation of
China. Z. F. L acknowledges financial support from the Research
Grant Council of Hong Kong through Project 402309. The computation is
performed in the Supercomputer Center of Shanghai and the
Supercomputer Center of ECNU.
\end{acknowledgments}

{\bf Supporting Information Available.} Two movies  for the permselectivity process of hydrogen in a mixture. This material is available free of charge via the Internet at http://pubs.acs.org.

\newpage

\begin{figure}[fig1]
\centering
\includegraphics[width=100mm]{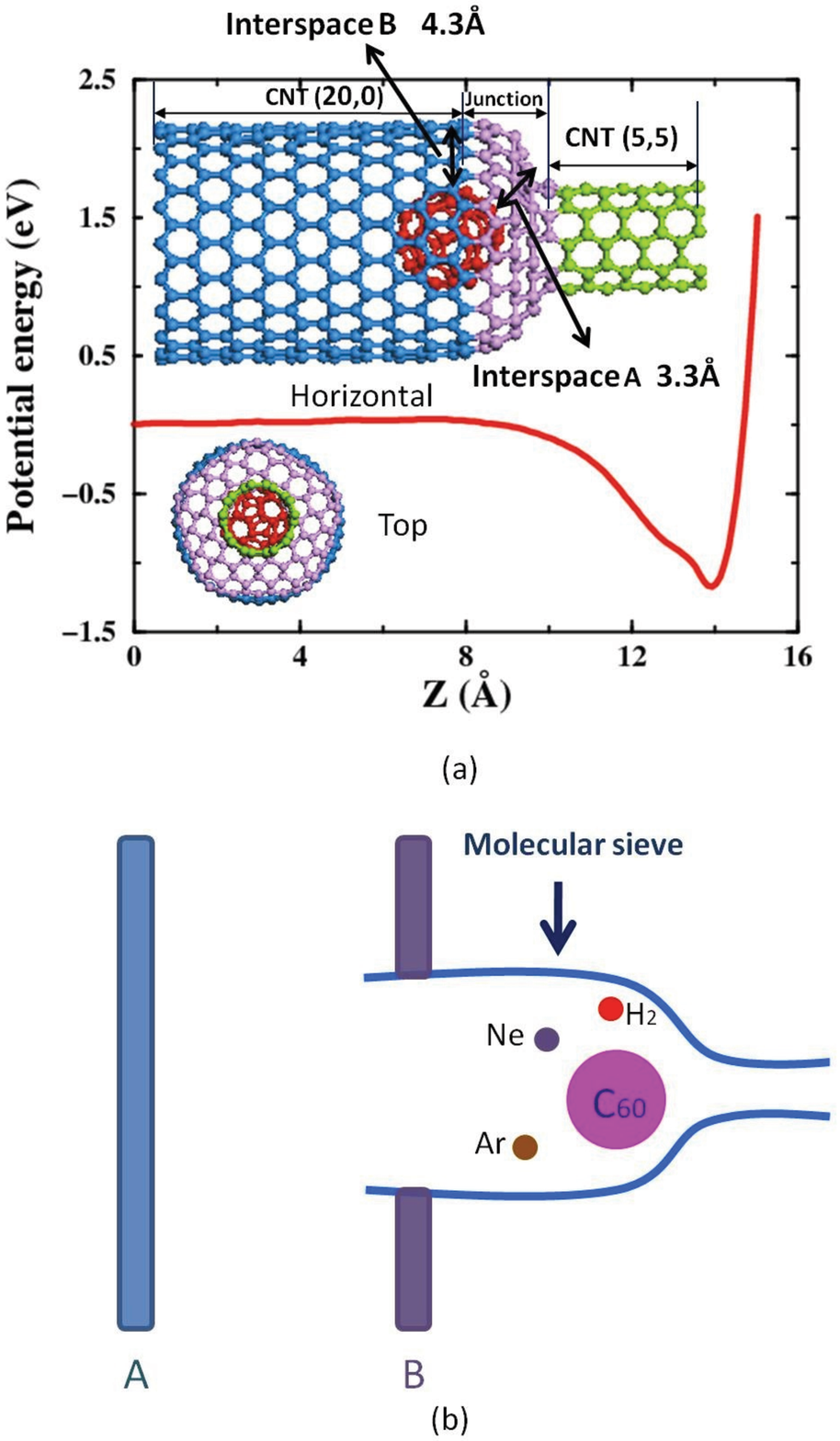}
\caption{(Color online) (a) The potential energy change of the
molecular sieve as $C_{60}$ approaching to the junction. The initial
axial distance of $C_{60}$ (the original of x-axis) is around
17$\AA$ far away from the junction. The energy has been shifted so
that the energy at initial position of $C_{60}$ is zero. Inset: The
top and horizontal views of the molecular sieve, which is
composed with a carbon nanotube intramolecular junction and an
encapsulated $C_{60}$. (b) The illustrations of the simulation
system for calculating the flux/permeability of gases, where
the plate B is fixed, while plate A is movable to
tune the pressure difference around the sieve.}
\end{figure}

\begin{figure}[fig2]
\centering
\includegraphics[width=100mm]{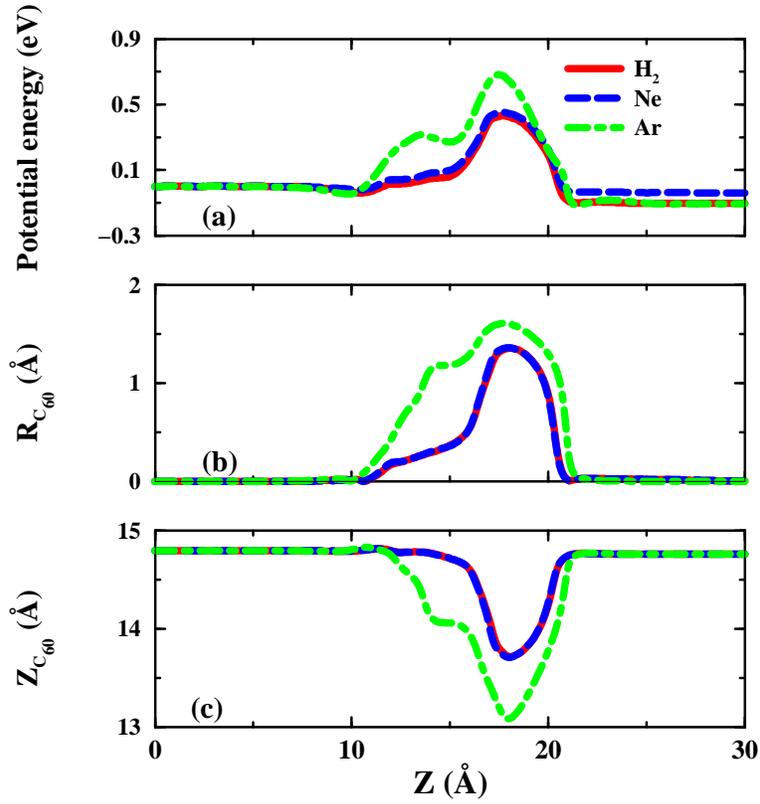}
\caption{(Color online) The potential energy change of system (a),
the radial (b) and axial displacement (c) of $C_{60}$ as the gas
molecule moves along axial direction. The initial axial distance of
molecule (the original of x-axis) is around 20$\AA$  far
away from the junction. The energy has been shifted so that the
energy at initial position of molecule is zero. }
\end{figure}

\begin{figure}[fig3]
\centering
\includegraphics[width=100mm]{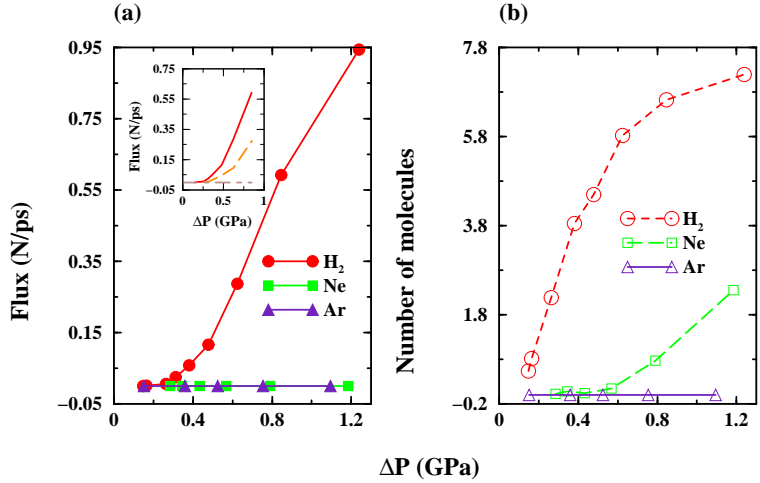}
\caption{(Color online) (a): The average flux varied as the function
of fluid pressure difference in the single component gases of $H_2$,
$Ne$ and $Ar$. Inset:  The flux for hydrogen in two different cases.
In the first case (long dashed line), the mass of $H_2$ is
artificially enlarged ten times. In the second (dot-dashed line),
the interaction of $H_2$-$H_2$ and $H_2$-C is  replaced by a typical
L-J potential. For comparing, the flux for hydrogen in normal
condition is also shown (solid line). (b): The average number of
molecules located in interspace B.  }
\end{figure}

\begin{figure}[fig4]
\centering
\includegraphics[width=100mm]{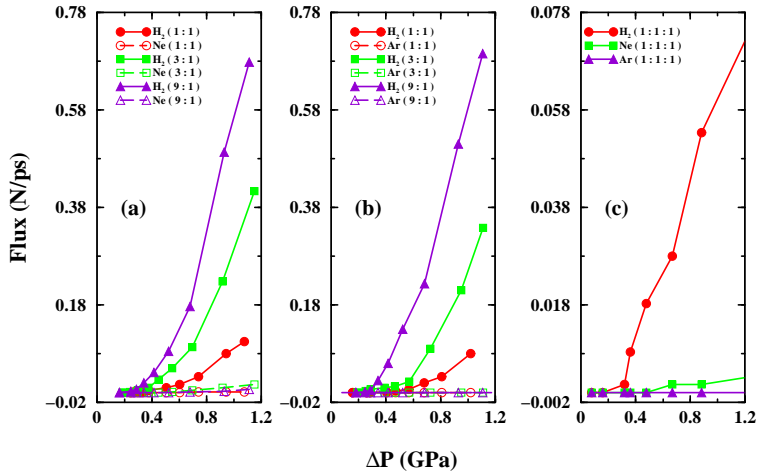}
\caption{(Color online) The average flux varied as the function of fluid
pressure difference in the multi-component gas molecules.(a) Mixed
$H_2$/$Ne$ with number ratio of 1:1/3:1/9:1. (b)Mixed
$H_2$/$Ar$ with number ratio of 1:1/3:1/9:1. (c)Mixed
$H_2$/$Ne$/$Ar$ with number ratio of 1:1:1.}
\end{figure}

\end{document}